\documentstyle[epsfig]{mn}

\def\kpc{{\rm\thinspace kpc}}

\def\kmps{\hbox{$\km\s^{-1}\,$}}
\def\cmps{\hbox{$\cm\s^{-1}\,$}}

\def\km{{\rm\thinspace km}}
\def\cm{{\rm\thinspace cm}}
\def\s{{\rm\thinspace s}}

\def\eV{{\rm\thinspace eV}}

\title[Radio galaxies and ICM X-ray absorption lines] {The influence
of radio-galaxy activity on X-ray absorption lines from the
intracluster medium}

\author[F.~K\"ockert \& C.~S.~Reynolds]{
\parbox{14cm}{Franziska~K\"ockert$^{1,2}$\footnotemark and Christopher~S.~Reynolds$^2$}\\
$^1$AIP, An der Sternwarte 16, D-14482 Potsdam, Germany.\\
$^2$Dept.\ of Astronomy, University of Maryland, College Park, MD 20742, USA.\\
}

\date{Accepted 2006 January 31. Recieved 2006 January 24; in original form
  2005 June 9}
\pagerange{\pageref{firstpage}--\pageref{lastpage}}
\pubyear{2006}

\begin{document}
\label{firstpage}

\maketitle

\begin{abstract}
 We present an investigation of the X-ray absorption features
 predicted by hydrodynamic simulations of radio galaxies interacting
 with the intracluster medium (ICM) of their host galaxy clusters.  We
 show how these absorption lines can be used as a new diagnostic for
 the radio-galaxy/ICM interactions.  Such interactions have been
 observed in numerous systems by {\it ROSAT}, {\it Chandra} and {\it
 XMM-Newton}, and understanding them has implications for AGN feedback
 and galaxy formation. Starting from the hydrodynamic simulations of
 Reynolds, Heinz \& Begelman, we calculate the properties of
 the highly ionized iron and oxygen lines (seen in absorption against
 the central active galactic nucleus; AGN), predicting line shapes,
 equivalent widths, column densities and velocity shifts. The main
 effect of the jet on the absorption lines is a reduction of the line
 strength from that of the quiescent ICM and the introduction of some
 velocity structure in the line profile. We investigate whether these
 features are detectable with current as well as future
 high-resolution X-ray spectrometers. We conclude that the {\it
 Chandra} transmission gratings have insufficient sensitivity to
 detect these features with high significance, and certainly would not
 allow a study of the dynamics of the interaction via absorption
 signatures.  {\it Constellation-X}, on the other hand, will allow
 superb constraints to be derived.  We can also use this analysis to
 assess the idea that radio-galaxy induced ICM outflows give rise to
 the resonant oxygen X-ray absorption lines that have been claimed as
 evidence for the warm-hot intergalactic medium (WHIM).  We show that
 these detached, high-velocity oxygen absorption lines cannot result
 from a radio-galaxy/ICM interaction, thereby strengthening the WHIM
 interpretation.
\end{abstract}


\footnotetext{E-mail: fkoeckert@aip.de (FK), chris@astro.umd.edu (CSR)}

\section{Introduction}

It has been known for over a decade that radio galaxies at the cores
of rich galaxy clusters interact with the surrounding intracluster
medium (ICM) and, in some cases, influence it in significant and
complex ways (e.g., B\"ohringer et al. 1993, 1995; Carilli, Perley \&
Harris 1994; Harris, Carilli \& Perley 1994; Heinz, Reynolds \&
Begelman 1998).  With the new generation of X-ray telescopes, {\it
Chandra} and {\it XMM-Newton}, observations of these interactions have
been more detailed than ever. For many systems it has been confirmed
that there are cavities in the ICM which show up as depressions in the
X-ray surface brightness and appear to be evacuated by the expanding
radio lobes (e.g., Hydra-A; McNamara et al. 2000, Abell~2052; Blanton
et al. 2001, Virgo-A; Young, Wilson \& Mundell 2002, Perseus-A; Fabian
et al. 2000, 2003a).  Old radio lobes which have remained intact
through magnetic or viscous effects (Reynolds et al. 2005, Kaiser et
al. 2005, Jones \& de~Young 2005) are also believed to be responsible
for the so-called ``ghost'' cavities seen in many clusters (e.g.,
Perseus, Fabian et al. 2003a; Abell~2597, McNamara et al. 2001;
Abell~4059, Heinz et al. 2002, Choi et al.  2004).  A second
indication of the importance of radio-galaxy/ICM interactions is the
cooling flow problem, that is the lack of cooled material in cluster
cores despite the fact that the radiative cooling time of many ICM
cores is rather short.  Observations by {\it XMM-Newton} have shown
that the gas does not in fact cool below 1--2\,keV (the so-called
temperature floor) begging us to identify a source of heating which
can offset the radiative cooling.

At the present time, the main observational diagnostics for
radio-galaxy/ICM interactions are the spatial mapping in the X-ray
band of temperature, pressure, cooling time and metal abundances of
the ICM across the interaction region.  However, with current data we
lack any direct probe of the dynamics of the interaction, that is the
velocities of the fluid disturbances, shocks and turbulence.  A major
hope for the {\it Suzaku} observatory was that it would open a direct
window on the dynamics of these interactions through the velocity
profiles of the ICM emission lines (Br\"uggen, Hoeft \& Ruszkowski
2005); of course, the {\it Suzaku} cryostat failure and subsequent
loss of the high-resolution X-ray Spectrometer (XRS) prevents this
goal from being realized.  Barring the emergence of a new mission
possessing a high-resolution X-ray spectrometer, this important
science goal must now await the launch of {\it Constellation-X}.

The current baseline design for {\it Constellation-X} has a
requirement on the spatial resolution of 15\,arcsec (half-power
diameter).  Coupled with its superior spectral resolution, this is
sufficient to allow detailed mapping of radio-galaxy driven ICM
dynamics through emission line profiles in the nearest systems such as
the {\it Virgo} and {\it Perseus} clusters.  For more distant systems,
however, the interaction region is contained within a small number of
spatial resolution elements, hampering the ability of the observation
to constrain robust dynamical signatures. In this paper, we discuss
the complementary technique of probing these radio-galaxy/ICM
interactions through absorption line spectroscopy of the X-ray
luminous core of the central radio galaxy, looking for jet-induced
changes to the strength and line profile of the absorption lines
present even in a static ICM.  While detection and characterization of
these absorption line features is more observationally demanding, it
has the major advantage of probing the kinematic state of the gas
through a well-defined ``core sample'' of the galaxy cluster, even if
ones X-ray observatory has only limited spatial resolution. We focus
on predictions of resonance K-shell absorption features of H- and
He-like oxygen and iron arising in the ICM around an active as well as
inactive radio galaxy, examining the velocity structure and strength
of the lines.

These radio-galaxy/ICM interactions are particularly interesting and
important since they appear to be the local and accessible examples of
the more general phenomena of ``AGN feedback'' and potentially crucial
for some aspects of cosmological structure formation (e.g., see Benson
et al. 2003 and references therein).  However, there is another
important reason to study this phenomenon; it is of direct relevance
to our observational understanding of the Warm-Hot Intergalactic
Medium (WHIM).  The filamentary WHIM is believed to be the repository
for half of the baryons in the local Universe and so is of obvious
interest to cosmologists. It is possible to detect the WHIM by wide
field imaging of diffuse EUV and soft X-ray emission (see for example 
Kaastra et al. 2003). However, the most important diagnostic of the
WHIM are UV and X-ray absorption lines of oxygen
seen in the spectra of bright AGN. Narrow oxygen OVI--OVIII absorption
lines at a redshift intermediate between zero and that of the AGN are
taken to be good candidates for the WHIM.  However, all AGN towards
which such lines have been claimed are radio-loud AGN (see Section 5).
The question arises whether these absorption line features are, in
fact, outflows associated with the jetted AGN.  While the possibility
remains that these features could be due to outflows associated with
the central engine itself, we demonstrate that it is not possible to
associate WHIM-like features with radio-galaxy/ICM interactions.

We predict the absorption line properties of radio-galaxy/ICM
interactions by using the XSTAR photoionization code to post-process
the hydrodynamics simulations of Reynolds, Heinz \& Begelman (2002).
Section 2 of this paper describes the underlying hydrodynamic
simulations of the radio galaxy we use. Section 3 explains the
simulation of our spectra using XSTAR.  In Section 4, we
summarize the properties (strengths and velocity structure) of the
absorption lines that we predict as a function of the inclination with
which we view the radio galaxy, the age of the radio-galaxy, and the
properties of the ICM into which the radio-galaxy is expanding.
Section 5 argues that these interactions cannot produce ``WHIM-like''
features, thereby strengthening the WHIM interpretation of these
intermediate redshift absorption lines.  Section 6 focuses on the
ability of current future X-ray observatories to study these
absorption lines and hence prove these radio-galaxy/ICM interactions.
Finally, Section 7 presents our conclusions.

\section{The underlying hydrodynamic simulations}

The hydrodynamic simulation underlying this work is the canonical
axisymmetric simulation of Reynolds, Heinz \& Begelman (2002); for
convenience, we briefly summarize the relevant details of the
simulation here.

This 2-d hydrodynamic simulation explores the evolution of an active
jet in a radio galaxy as it interacts with a surrounding ICM, and the
resulting structure after the jet is shut down.  Initially, we start
with an isothermal spherical gaseous atmosphere with a (number)
density profile given by
\begin{equation}
n(r)=\frac{n_0}{\left[1+(r/r_0)^2\right]^{3/4}}.
\end{equation}
The (fixed) gravitational potential is defined such that this
atmosphere is in hydrostatic equilibrium.  Back-to-back supersonic
jets are injected in initial pressure balance with a density of
$n_{\rm jet}=n_0/100$ and Mach number (with respect to the internal
jet sound speed) of 10.  The half-opening angle of the jet is
$15^{\circ}$, and the starting point for the jets is a radius of
$r_0/20$. As discussed in the Appendix of Reynolds, Heinz \& Begelman
(2002), this choice of parameters results in a cocoon in which the
Kelvin-Helmholtz growth rates approximately match those calculated for
real systems.  Code units are related to physical units by parameters
fixed to the background medium. For most runs we use the following
canonical conversion factors: $r_{0}=100\,{\rm kpc}$, $c_{\rm
ISM}=1000\,{\rm km}\,{\rm s}^{-1}$, $n_{0}=0.01\,{\rm cm}^{-3}$ and
$T_{0}=4.4\times 10^{7}$ K. This gives a total kinetic luminosity of
the jets of $9.3\times 10^{45}\,{\rm erg}\,{\rm s}^{-1}$, making the
simulation relevant for powerful radio sources located in rich galaxy
clusters, like Cygnus A. We will refer to this as our ``large cluster
case''.  Since many radio galaxies are, however, found in smaller and
cooler galaxy clusters, we also consider a different scaling
($r_{0}=10\,{\rm kpc}$, $c_{ISM}=500\,{\rm km}\,{\rm s}^{-1}$,
$n_{0}=0.05\,{\rm cm}^{-3}$ and $T_{0}=1.1\times 10^{7}$ K) which
results in a total kinetic source luminosity of $1.16\times
10^{44}\,{\rm erg}\,{\rm s}^{-1}$ and is closer to objects like Hydra
A and Virgo A. We will refer to this as our ``small cluster case''.
In the first case, one code unit of time corresponds to 50 Myr. The
total length of the simulation is 240 Myr, the jet is turned off after
50 Myr. In the second case, one code unit of time is 10 Myr, the total
length is about 50 Myr, and the jet is turned off after 10 Myr. For
more detailed information on the simulations, see Reynolds, Heinz \&
Begelman (2002). As an example of the simulation results, a density
map is given in Fig.~1.

It is important to note that since this simulations were performed,
Chandra observations have revealed several cases where the radio lobes
of the central source are surrounded by a significant amount of cold
gas.  The ICM structures surrounding Perseus-A provide an excellent
example (see Fabian et al. 2003a; Schmidt, Fabian \& Sanders 2002). At
the current time, the physical mechanisms underlying these cool rims
are not understood and they cannot be reproduced in any first
principles simulations.  In systems containing such rims, the observed
oxygen OVII and OVIII absorption lines towards the central AGN are
likely to be dominated by the contribution from gas in the rim;
according to Schmidt, Fabian \& Sanders (2002), the temperature and
density in the cold gas rim of the Perseus cluster are
$3.6\times10^{7}$ K and $0.12\,{\rm cm}^{-3}$ respectively. Using
simple calculations based on Shull \& van Steenberg (1982), this gives
an OVIII equivalent width of 0.15\,eV, larger than that predicted from
the entire rest of the ICM atmosphere.  Therefore, our results
presented here are valid only for systems that do not possess these
cold rims, or for lines of sight that do not intercept such rims.

\begin{figure}
\hbox{
\psfig{figure=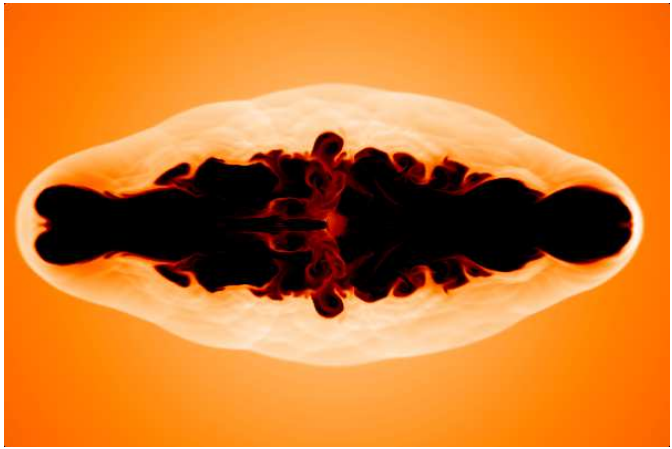,width=0.5\textwidth}
}
\caption{Density map of the jet and its close environment.  In our
large-cluster (``Cygnus-A'') scaling , the jet in this snapshot has
been active for 30\,Myr and the region displayed is the central
$200\kpc\times 300\kpc$ of the cluster.}
\end{figure}

\section{Simulation of the absorption features}

The hydrodynamic simulation gives us the density, temperature and
velocity of the ICM at every point in space as a function of time
throughout the modelled radio-galaxy/ICM interaction.  To simulate the
absorption line spectrum that one would see towards the central radio
galaxy, we used the photoionization code XSTAR (developed by Tim
Kallman; see Kallman \& Bautista 2001) to calculate the opacity of the
ICM along radial rays that originate at the AGN.  We assume a
collisionally-ionized plasma in thermal and ionization equilibrium.
These assumptions are readily verified for the ions of interest.

We take the temperature, density and velocity directly from the
hydrodynamic simulation.  The frequency-dependent opacity is
calculated including both thermal line broadening and line-of-sight
Doppler shifts.  We choose an energy resolution of $\Delta E/E =
3.29\times 10^{-4}$, giving a velocity resolution of about 100 km/s.

XSTAR includes 13 elements, and the elemental abundances are important
for the resulting line spectrum.  We chose abundances measured by
Allen \& Fabian (1998), Loewenstein \& Mushotzky (1996) and
Baumgartner et al. (2005), which we summarize in Table 1.

\begin{table*}
\begin{tabular} {llllllllllll}
\hline
\hline
Element&C&N&O&Ne&Mg&Si&S&Ar&Ca&Fe&Ni\\
$Z/Z_\odot$&$0.34^{\rm a}$&$0.34^{\rm a}$&$0.48^{\rm b}$&$0.62^{\rm b}$&$0.376^{\rm b}$&$0.63^{\rm c}$&$0.26^{\rm c}$&$0.17^{\rm c}$&$0.275^{\rm c}$&$0.401^{\rm c}$&$1.02^{\rm c}$\\
\hline
\end{tabular}
\caption{Cosmic element abundances compared to solar.  $^{\rm a}$Allen
\& Fabian 1998, $^{\rm b}$Loewenstein \& Mushotzky 1996, $^{\rm
c}$Baumgartner et al. 2005}
\end{table*} 

All publicly available photoionization codes have some problems
dealing with very low densities and very high temperatures ($>10^{9}
K$); XSTAR is no exception. However, such high temperatures only occur
in early times of the simulations, when the jet is still on, and then 
only in the innermost grid cells, very close to the source. We can
safely ignore those cells (through the inclusion of a threshold
temperature above which the simulation cell is skipped) since very
high temperature gas has no line absorption anyway.  The only case
where a considerable number of grid cells is affected is along the
jet-axis $\theta<1^\circ$.  Due to the forced symmetry of the
simulations the region $\theta<1^\circ$ is unphysical because material
gets ``stacked'' on the jet axis.  The minimum line-of-sight angle
that we choose to analyze is $\theta=5.5^\circ$, appropriate for
blazar-type sources.

\section{Results}

In this section, we discuss the impact that the radio-galaxy activity
has on the observed X-ray absorption lines in our simulated system.
The astrophysical relevance of these results, and the ability of
current and future instruments to measure the predicted absorption
lines will be addressed in the following sections.

We performed two sets of absorption line simulations. The first set
examines different lines of sight towards the AGN, at four chosen
times of the jet evolution.  Using the large cluster case, we chose to
examine the spectrum of the system at 30\,Myr (when the jet is fully
active), at 50\,Myr (just when the jet is turned off), 100\,Myr (well
after it is turned off) and at the end of the simulation at
240\,Myr. Our second set of calculations looks at the continuous
evolution in time, along five representative lines of sight which make
an angle $\theta$ = 5.5, 12, 30, 60 and 90 degrees with the jet axis.

While our calculations actually include the thousands of lines that
are incorporated into XSTAR, we focus our examination on the K-shell
resonant absorption lines of FeXXV (6.70\,keV), OVII (0.574\,keV) and
OVIII (0.654\,keV) lines since these are expected to be the strongest
and/or most interesting. After examining the possibility of observing
the features with {\it Constellation-X}, we decided to also carefully
examine the SiXIII line (1.86\,keV) since it turned out to be
particularly strong and at an energy that is particularly well-tuned
to {\it Constellation-X}.  We note that the FeXXVI lines (6.95\,keV
and 6.97\,keV) will not be discussed in much detail since they prove
too weak to be detected even with {\it Constellation-X}. For all
cases, we examine the optical depth, equivalent width, column density
and velocity shifts. Figure 2 shows an example for the simulated full
spectrum for $\theta=5.5^{\circ}$, t=50 Myr (large cluster
case). Figure 3 shows the strengths of the most prominent lines in
this case.

\begin{figure}
\hbox{
\psfig{figure=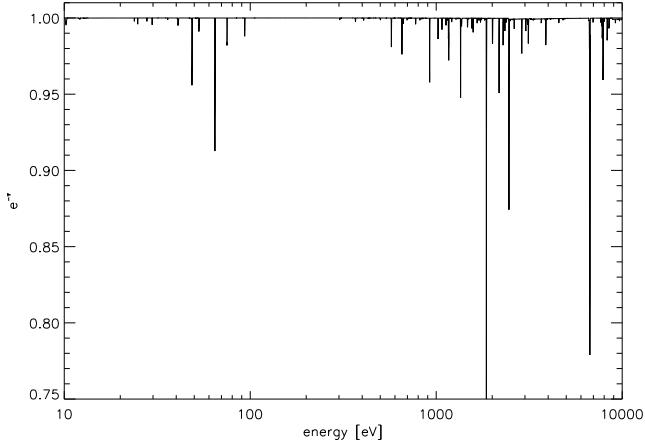,width=0.5\textwidth}
}
\caption{The simulated XSTAR spectrum for $\theta=5.5^{\circ}$ and
t=50 Myr (large-cluster case).\label{fig2}}
\end{figure}

\begin{figure}
\hbox{
\psfig{figure=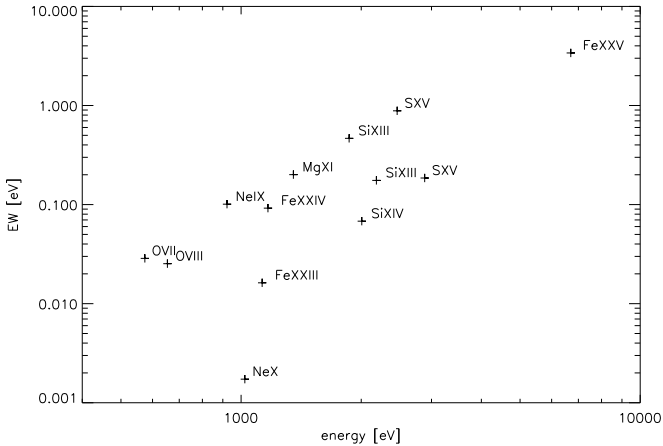,width=0.5\textwidth}
}
\caption{The strength of the most prominent lines in the simulated
  spectrum for $\theta=5.5^{\circ}$ and $t=$50\,Myr (large-cluster
  case).\label{fig3}}
\end{figure}

\subsection{The undisturbed material}

Even the undisturbed ICM produces significant oxygen and iron line
absorption (Krolik \& Raymond 1988; Sarazin 1989).  Undisturbed lines
arise from material which is either ahead of the shock front, or which
has settled down again since enough time has passed since the jet was
turned off. For our large cluster case, the FeXXV and FeXXVI
equivalent widths are 11.65\,eV and 0.086\,eV (corresponding to column
densities of $N_{\rm FeXXV}=1.32\times10^{17}\,{\rm cm}^{-2}$ and
$N_{\rm FeXXVI}=2.81\times10^{15}\,{\rm cm}^{-2}$) whereas for OVII,
OVIII and SiXIII we predict 0.05\,eV, 0.155\,eV and 1.536\,eV
(corresponding to column densities of
$N_{OVII}=6.46\times10^{14}\,{\rm cm}^{-2}$,
$N_{OVIII}=1.68\times10^{15}\,{\rm cm}^{-2}$ and
$N_{SiXIII}=1.84\times10^{16}\,{\rm cm}^{-2}$).

For the small cluster case, there is no FeXXVI line since the ICM does
not get hot enough for appreciable H-like iron to exist. FeXXV and
SiXIII have equivalent widths of 7.64\,eV and 11.97\,eV (corresponding
to column densities of $N_{\rm FeXXV}=8.67\times10^{16}\,{\rm
cm}^{-2}$ and $N_{SiXIII}=1.43\times10^{17}\,{\rm cm}^{-2}$), OVIII
and OVII have equivalent widths of 1.004\,eV and 0.548\,eV
(corresponding to column densities of
$N_{OVIII}=1.1\times10^{16}\,{\rm cm}^{-2}$ and
$N_{OVII}=7.12\times10^{15}\,{\rm cm}^{-2}$). The oxygen and SiXIII
lines are stronger than in the large-cluster case since there is more
cooler material to contribute to the absorption, but the FeXXV line is
weaker due to the low assumed temperature of the ICM.

\subsection{Disturbances introduced by the jet}

When the AGN jet moves through the ICM, it influences it in numerous
ways. First, it can induce motions in large amounts of material. As
the ICM gets swept up in the shockfront, some of it moves outwards
with the shocked shell and some gets caught in a backflow. In the late
stages of the evolution of the system, a significant amount of ICM
falls back into the core regions as the cocoon collapses and buoyantly
rises into the cluster atmosphere.  All of these effects will imprint
signatures (broadening, and the development of blueshifted and
redshifted wings and peaks) in the absorption line profiles. The
second strong effect is a change of the ICM temperature. Strong shocks
associated with the early phase of jet activity will appreciably heat
some regions of the ICM, typically raising it to temperatures
sufficient to fully ionize oxygen.  On the other hand, unshocked ICM
that is caught in the ``updraft'' from a buoyant cocoon will
adiabatically decompress and be cooled (strengthening oxygen
absorption features). 

Clearly, the kinematics and the thermodynamics are coupled.  We might
expect (and, as discussed below, confirm) it to be appreciably harder
to form high velocity oxygen absorbers than iron absorbers, since any
strong shock responsible for accelerating the gas will inevitably heat
it to the point where oxygen becomes fully ionized.  On the other
hand, we find iron lines which, in the strongest cases, display a
double peak structure, with one peak being at the rest energy of the
line and one being clearly blueshifted but still attached to the line
itself. In those cases, the blueshifted components arise from a large
amount of shock accelerated material outflowing at fairly high speeds,
whereas the peak at the rest energy comes from still undisturbed
material in front of the jet.  

Representative results for oxygen and iron line profiles and
equivalent widths are shown in Figs.~4--7.  It is instructive to
separate the discussion of the {\it cocoon inflation} phase of
activity from that of the {\it cocoon collapse and buoyant plume
formation} phase.  During the cocoon inflation phase, the radio galaxy
activity inflates an over-pressured cocoon which, in turn, drives a
strong shock into the ICM.  The high ICM temperatures produced by this
shock leads to a rapid drop in the equivalent widths of the OVII,
OVIII, SiXIII and FeXXV lines with rather little change in the line
profile.  The decrease in line equivalent widths are rather more
dramatic for lines of sight close to the jet axis simply due to the
increased path-length along which the ambient line-absorbing ICM has
been scoured out by the cocoon shock.

Towards the end of the cocoon-inflation phase, the cocoon pressure
becomes comparable to that of the ambient ICM and the shock weakens.
After that, the source transits into the cocoon-collapse phase in
which the shocked ICM surrounding the sides of the cocoon (i.e., the
equatorial regions with respect to the jet axis) falls back towards
the cluster center.  The result of this infall is to ``squeeze'' the
cocoon plasma and transform it into two buoyantly rising,
mushroom-shaped plumes (see Fig.~1 of Reynolds, Heinz \& Begelman
2002).  During these times, the equivalent widths of the OVII, OVIII,
SiXIII and FeXXV absorption lines gradually recover to approximately
their initial values.  However, the complex dynamics of the ICM during
this phase create imprints in the line profiles.  For lines of sight
close to the jet axis, both the oxygen and iron absorption lines
develop blue wings corresponding to absorption by the outward moving
ICM shell and, latter, ICM that is being dragged out of the core
regions of the cluster in the wake of the buoyant ICM plume.  These
features are subtle in the case of the oxygen lines, but are rather
dramatic in the case of the FeXXV lines.  In all cases, the velocities
characterizing the blue-wing are less than but of the order of the ICM
sound speed.  Interestingly, for lines of sight that make a large
angle with the jet-axis, the FeXXV line displays a subtle redshift of
its centroid corresponding to the actual inward collapse of the ICM
core (Fig.~7).

\begin{figure*}
\hbox{
\psfig{figure=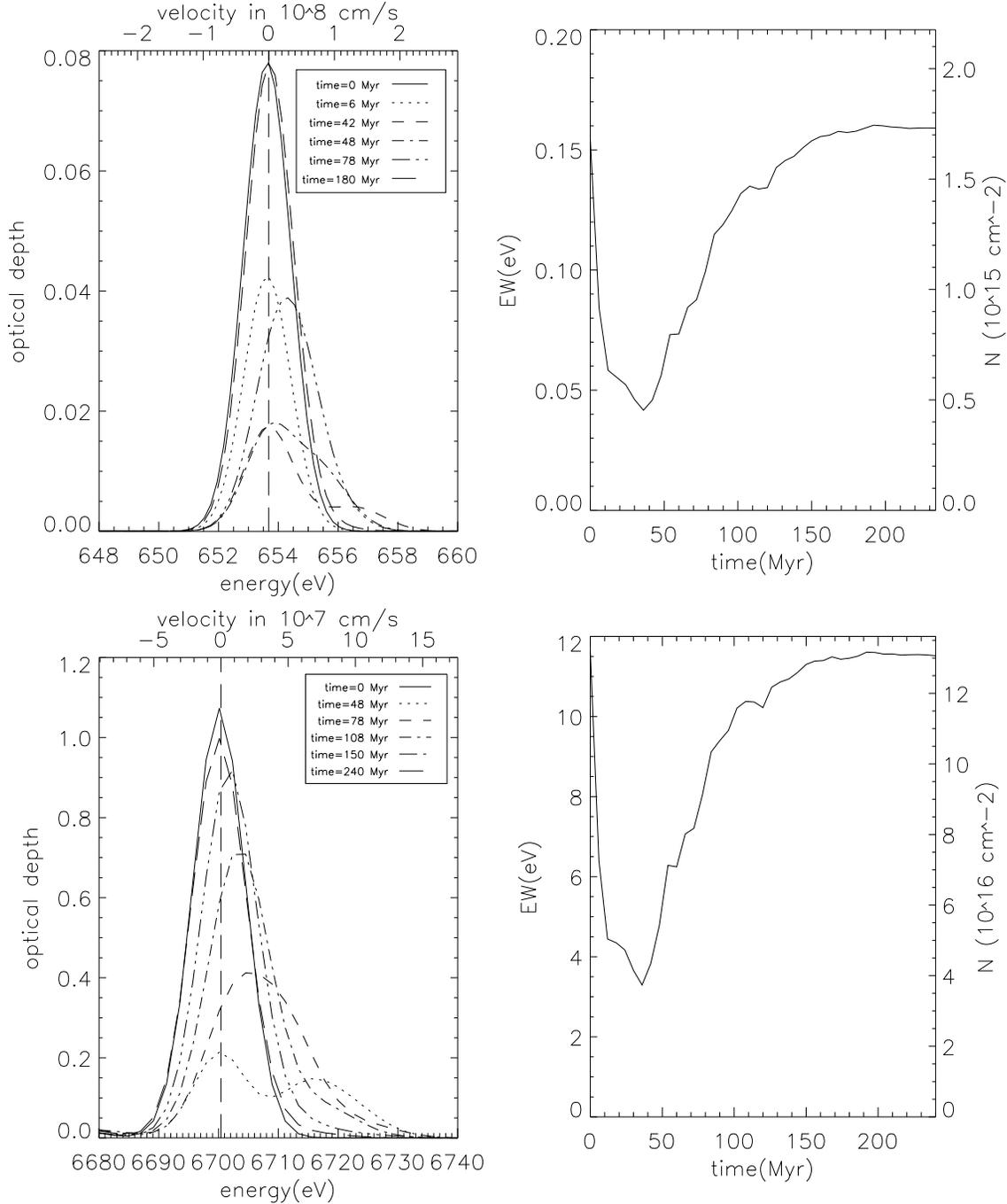,width=0.9\textwidth}
}
\caption{Line profiles and equivalent widths as a function of time for
  the small inclination case of $\theta=5.5^\circ$ and the
  large-cluster scaling.  {\it Upper left panel: }Energy profiles of
  the OVIII K$\alpha$ resonance absorption line.  {\it Upper right
  panel : }Total equivalent width of the OVIII K$\alpha$ resonance
  absorption line as a function of time.  {\it Lower left panel:
  }Energy profile of the FeXXV K$\alpha$ resonance absorption line.
  {\it Lower right panel: }Total equivalent width of the FeXXV
  absorption lines as a function of time.  Note the growth of an
  extended blue-shifted wing at the end of the cocoon-inflation
  phase.}
\end{figure*}

\begin{figure*}
\hbox{
\psfig{figure=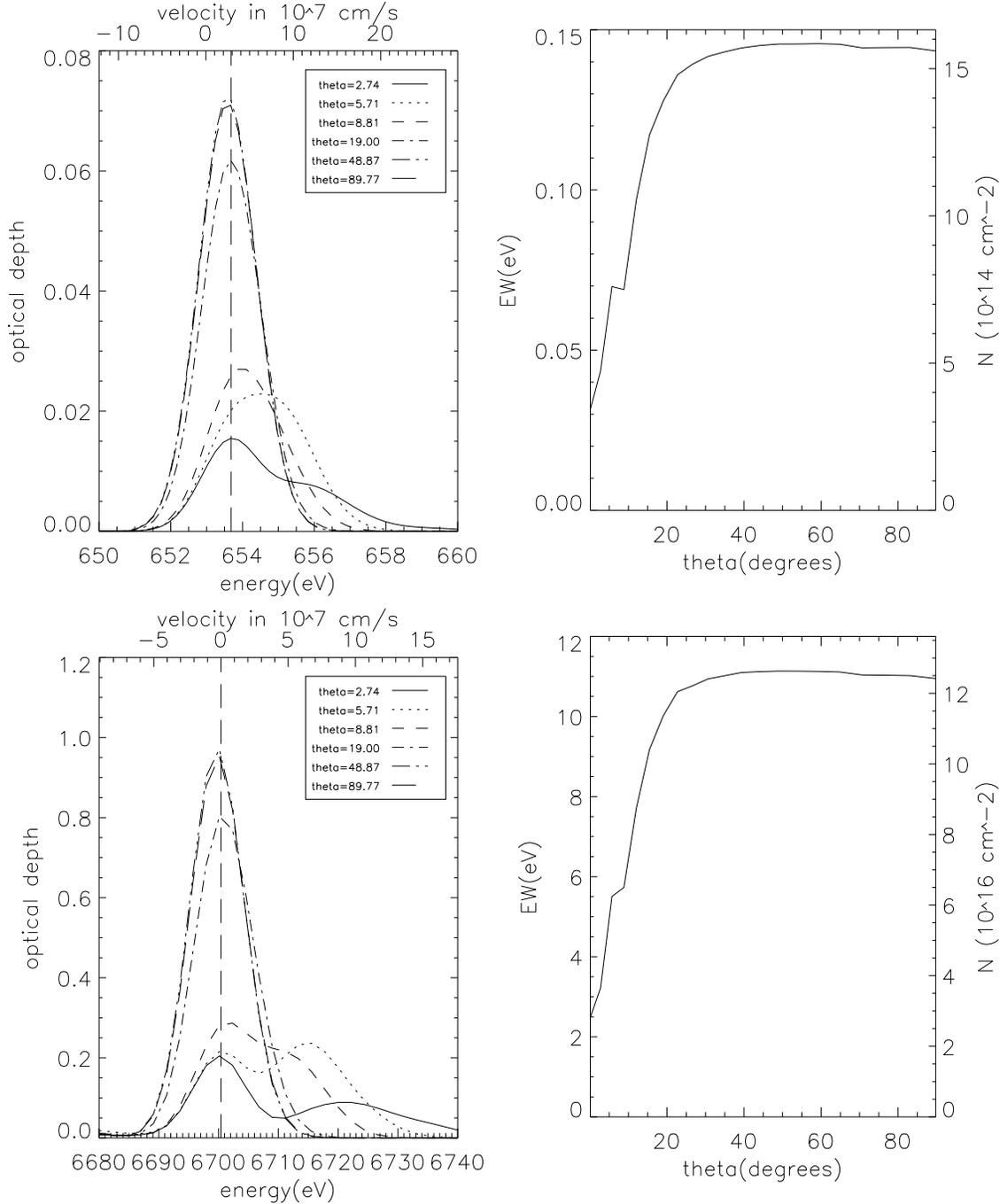,width=0.9\textwidth}
}
\caption{Line profiles and equivalent widths as a function of viewing
  angle at the time that the jet shuts off ($t=50$\,Myr) for the
  large-cluster scaling.  {\it Upper left panel: }Energy profiles of
  the OVIII K$\alpha$ resonance absorption line.  {\it Upper right
  panel : }Total equivalent width of the OVIII K$\alpha$ resonance
  absorption line as a function of inclination.  {\it Lower left
  panel: }Energy profile of the FeXXV K$\alpha$ resonance absorption
  line.  {\it Lower right panel: }Total equivalent width of the FeXXV
  absorption lines as a function of inclination.  Note the growth of
  an extended blue-shifted wing at small inclinations.}
\end{figure*}

\begin{figure*}
\hbox{
\psfig{figure=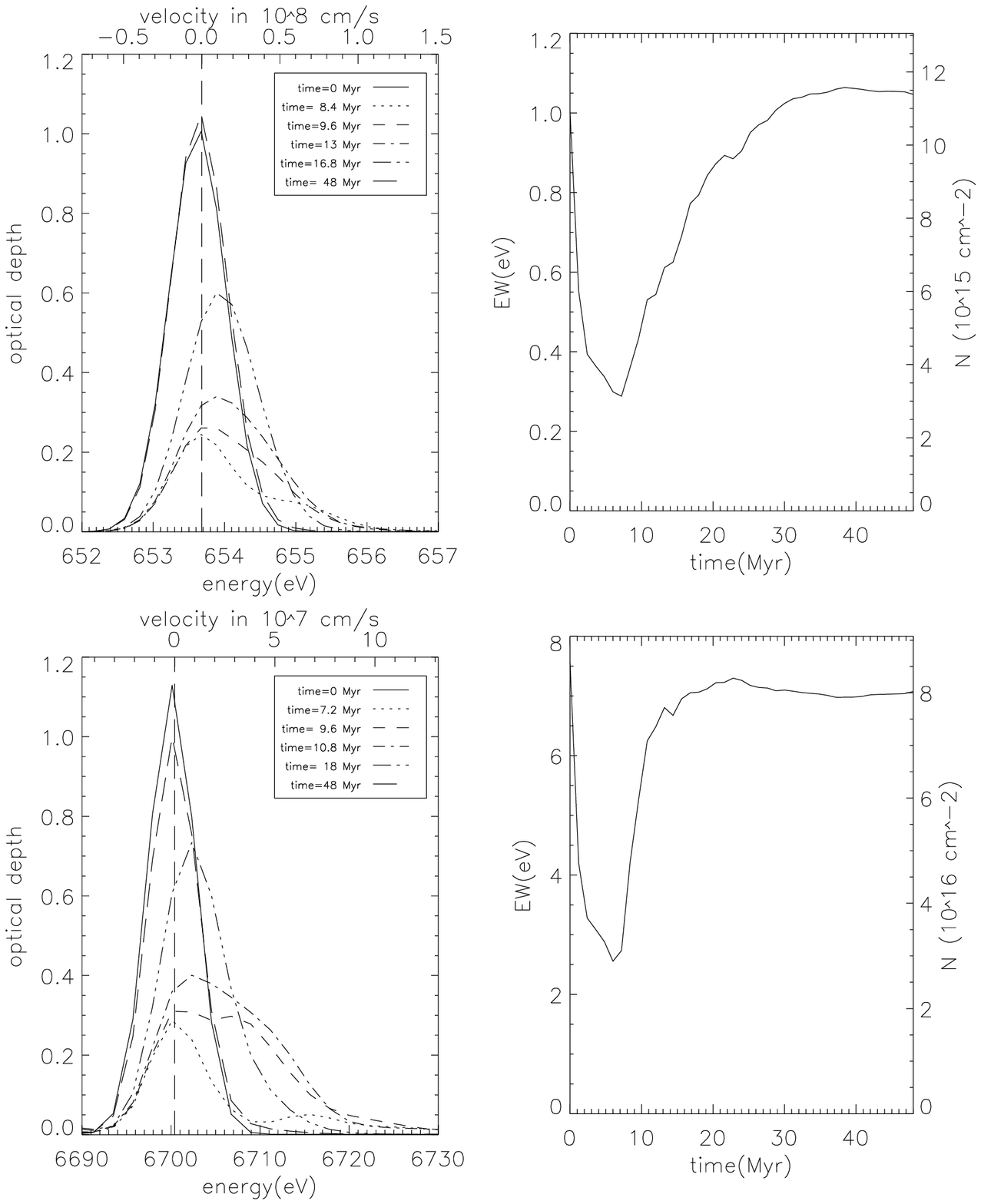,width=0.9\textwidth}
}
\caption{Line profiles and equivalent widths as a function of time for
  the small inclination case of $\theta=5.5^\circ$ and the
  small-cluster scaling.  {\it Upper left panel: }Energy profiles of
  the OVIII K$\alpha$ resonance absorption line.  {\it Upper right
  panel : }Total equivalent width of the OVIII K$\alpha$ resonance
  absorption line as a function of time.  {\it Lower left panel:
  }Energy profile of the FeXXV K$\alpha$ resonance absorption line.
  {\it Lower right panel: }Total equivalent width of the FeXXV
  absorption lines as a function of time.  Note the growth of an
  extended blue-shifted wing at the end of the cocoon-inflation
  phase.}
\end{figure*}

\begin{figure*}
\hbox{
\psfig{figure=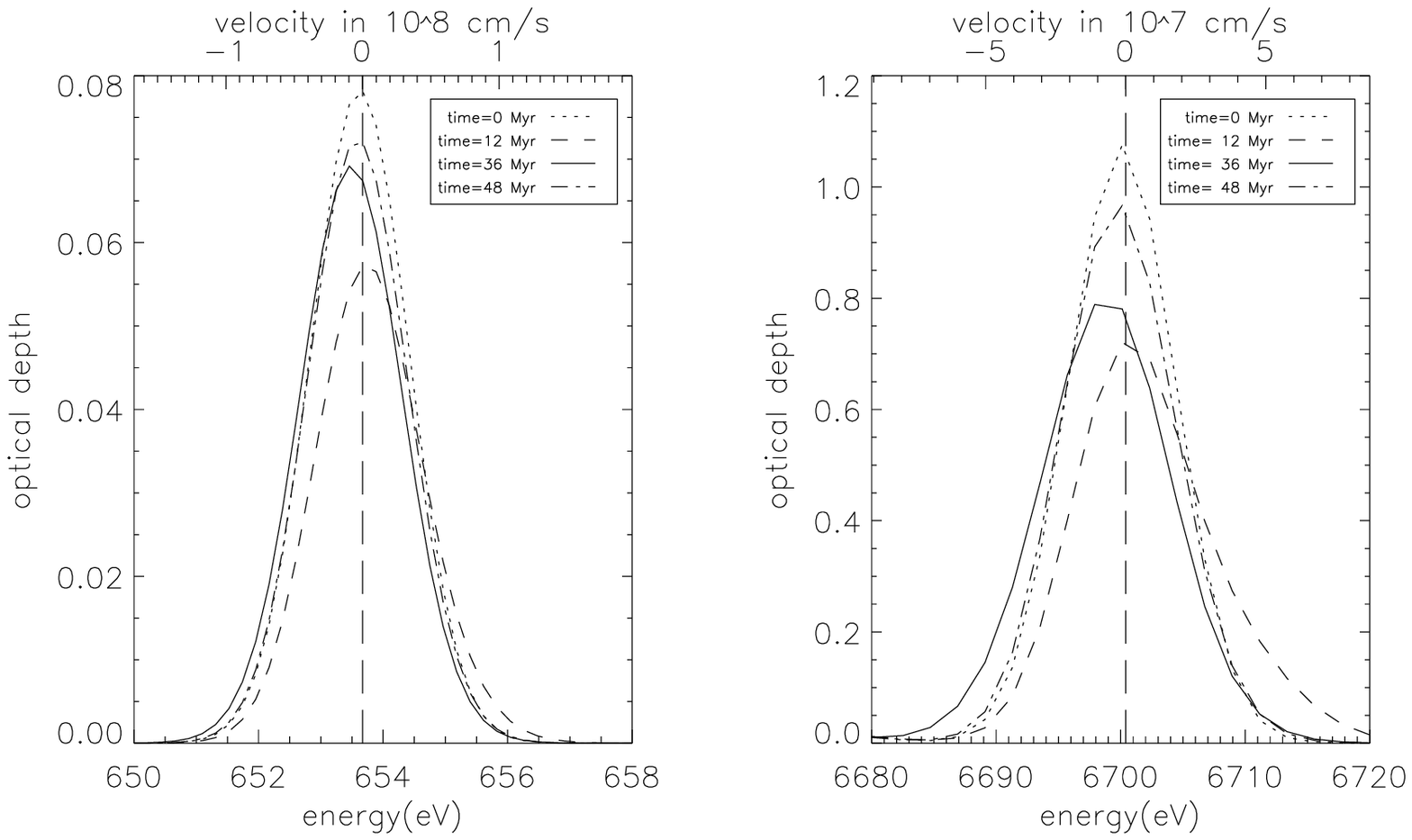,width=0.9\textwidth}
}
\caption{Line profiles and equivalent widths as a function of time at 
  a viewing angle of $60^{\circ}$ for the large-cluster scaling. {\it Left
  panel: }Energy profiles of the OVIII K$\alpha$ resonance absorption
  line.{\it Right panel: }Energy profile of the FeXXV K$\alpha$
  resonance absorption line. Note the subtle redshift of the centroid of both
  lines associated with the lateral collapse of the cocoon, at t=36 Myr (solid
  line).}
\end{figure*}

\section{Implications for the WHIM observations}

There has been much recent excitement about the possible detection of
the WHIM through the OVII and OVIII K-shell resonant absorption lines.
This is of obvious importance given that these WHIM filaments are
expected to be the repository for half of the baryons in the local
Universe.  The most robust detection to date was obtained by a {\it
Chandra} High Energy Transmission Gratings (HETG) observation of the
blazar Mrk~421 during an outburst (Nicastro et al. 2005a,b).  Two
absorber systems were detected in both OVII and NVII at velocities of
$cz=3300\pm 300\kmps$ and $cz=8090\pm 300$ (to be compared with the
recession velocity of the blazar of $CZ=9000\kmps$).  The OVII
K$\alpha$ EWs of these two systems were measured to be $0.080\pm
0.021\eV$ ($3.0\pm 0.8 m\AA$) and $0.059\pm 0.021\eV$ ($2.2\pm 0.8
m\AA$).  Nicastro et al. (2005b) demonstrate that the column density
distribution implied by these detections is consistent with the notion
that the WHIM filaments do, indeed, balance the baryon budget of the
local Universe.  While this is the most robust detection of $z>0$
WHIM, it was not the first.  Fang et al. (2002) claim a detection of
the OVIII K$\alpha$ line with an EW of $0.41\eV$ from a system with
$cz=16,600\kmps$ towards the blazar PKS~2155--304 (which has a
systemic velocity of $cz=34,800\kmps$), although the significance of
this detection (which is based on finding a single line in a blind
search of the spectrum) has been questioned.  McKernan et al. (2003)
also claim a OVIII line with a velocity of $cz=4400\kmps$ and a EW of
$2.17\eV$ towards the broad line radio galaxy 3C120 (which has a
systemic velocity of $cz=9900\kmps$).  Other recent detections of
$z>0$ WHIM via X-ray absorption lines have been reported towards the
BL-Lac object PKS~0548--322 (Barcons et al. 2005) and the radio-loud
quasar H1821+643 (Mathur, Weinberg \& Chen 2003).

While observational biases are evident, it is striking that all of the
claimed detections of $z>0$ WHIM are along lines of sight towards
radio-loud AGN.  The question arises as to whether some aspect of
jetted AGN can produce high-velocity and detached X-ray absorption
lines.  With our calculations, we can address one particular class of
models, i.e., those where the high-velocity X-ray absorption lines are
produced in an jet-induced ICM outflow.  In particular, we search for
phases of activity in which one would see narrow and detached oxygen
absorption lines.  In fact, at no point in time do we ever see such
features.  Instead, any high velocity material always appears as a
blueshifted tail on the main zero velocity absorption line.
Furthermore, our velocity shifts are on the order of the sound speed in
our ICM.  Any ICM that is accelerated to appreciably higher velocities
will be shock heated to a point where it can no longer produce oxygen
absorption lines (i.e., the oxygen will be fully ionized).  

Thus, we conclude that the observed WHIM-like absorption line systems
cannot arise from ICM/jet interactions.  This leaves two
possibilities.  One would be nuclear absorption, meaning absorption
from outflowing material in the central nucleus of the AGN, for
example from an accretion disk wind. This could be tested by looking
for variability of the line, since the region of origin would be
comparatively small. The other possible explanation remains the WHIM.
It is important to obtain deep high-resolution spectra of radio-quiet
AGN in order to assess this possibility.

\section{Observing absorption lines resulting from ICM-RG interactions}

We have used the XSPEC spectral fitting package to investigate the
possibility of observing our simulated absorption features with {\it
Chandra}, {\it Suzaku} and {\it Constellation-X}.  To simulate ICM
absorption towards a bright radio-galaxy, we use our calculations to
predict the absorption towards a power-law source with photon index
$\Gamma=1.6$ and flux $3\times10^{-11}\,{\rm erg\,cm^{-2}\,s^{-1}}$.
These parameters are choosen to mimic the radio-galaxy 3C~120
(McKernan et al. 2003).  Note that while our hydrodynamic simulations
were motivated by Cygnus-A, the highly absorbed nature of the nucleus
of Cygnus-A makes it inappropriate as a target for such absorption
line studies.  Given a predicted spectrum, we use the ``fakeit'' tool
in XSPEC to simulate spectra using the appropriate response matrices
for the high-energy transmission gratings (HETG) on {\it Chandra}, the
now in-operable X-ray spectrometer (XRS) on {\it Suzaku} and the
microcalorimeter on {\it Constellation-X}.  We chose the optimistic
case of integration times of 200\,ks.

To determine the detectability of the lines of interest, we first
fitted the simulated data with a powerlaw.  We then search for the
absorption lines by adding a simple negative Gaussian model and
applying the F-test to the resulting improvement in the goodness of
fit.  This procedure reveals that neither {\it Chandra}/HETG nor {\it
Suzaku}/XRS are able to detect the iron and oxygen absorption features
for lines-of-sight close to the jet axis, because along those
lines-of-sight the predicted features are also the weakest. The only
detectable feature at $\theta=5.5^{\circ}$ is the SiXIII line. The
simulated line as detected by {\it Suzaku} is shown in
Fig.~8. Detections of the other features with {\it Suzaku} are
possible at higher inclinations.  As discussed in Section 4 however,
the ICM dynamics imprint only subtle features on the line profile for
these higher inclinations.  Thus, we conclude that studies of ICM
dynamics using these absorption lines are not possible with the
current generation of instruments.

On the other hand, NASA's proposed {\it Constellation-X} observatory
will have such improved sensitivity that it will easily allow the
study of jet-driven ICM dynamics using absorption line profiles.
Looking at the case of $\theta=5.5^{\circ}$ and t=50 Myr, the most
prominent lines picked up by {\it Constellation-X} are SiXIII, FeXXV,
SXV, MgXI, FeXXIV, NeIX, OVIII, NeX and OVII. As shown in Fig.~9,
FeXXV displays the double peaked structure with one strongly
blueshifted second peak introduced by the outflowing shocked ICM
shell.  SiXIII and SXV both display blueshift wings making a Gaussian
not an adequate model anymore. All lines display blueshifted peaks
with $\Delta v=10^{6}$ to $10^{7}\cmps$. The simulated FeXXV line
detection is shown in Fig.~9.

\begin{figure}
\hbox{
\psfig{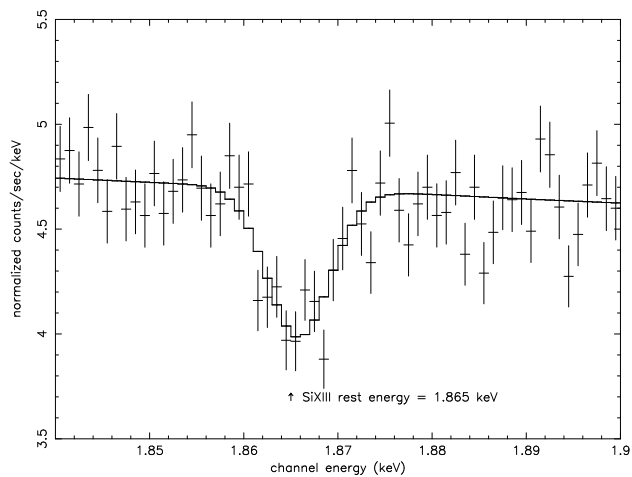}
}
\caption{Simulated {\it Suzaku} detection of the SiXIII line in the
  large-cluster case for an inclination of $\theta=5.5^{\circ}$, and a
  time of $t=50$\,Myr.  See text for details of the simulated source
  and observation.}
\end{figure}

\begin{figure}
\hbox{
\psfig{figure=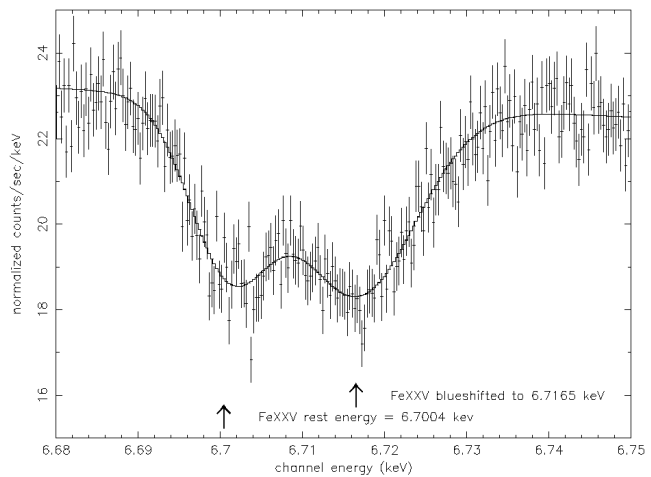,width=0.5\textwidth,angle=-90,scale=0.7}
}
\caption{Simulated {\it Constellation-X} detection of FeXXV in the
  large-cluster case for an inclination of $\theta=5.5^{\circ}$, a
  time of $t=50$\,Myr.  See text for other details of the simulated
  source.  Note the double-peak structure.  The blueshifted peak
  corresponds to material in the approaching shocked-shell of ICM.}
\end{figure}

\section{Conclusions}

In this paper we present spectral simulations of ICM absorption
features arising from radio-galaxy/ICM interactions. These features
can be used as a new tool to probe interactions between the ICM and
radio galaxies. The energy introduced by the jet weakens the lines
already present due to the static ISM and, more importantly,
introduces a velocity structure in the form of wings and blueshifted
line peaks in the line profile.  We specifically looked at OVIII, OVII
and FeXXV. The results were quantified as equivalent widths, column
densities and velocity shifts of those absorption lines. We used the
results for the oxygen lines to test the possibility of jet-induced
ICM outflow giving rise to absorption lines which could be mistaken as WHIM 
in the claimed observations. Our
results largely rule that out and strengthen the WHIM
interpretation. We also tested the possibility of actually observing
those features introduced by the jet with the HETG on {\it Chandra},
the now inoperable XRS on {\it Suzaku} and the microcalorimeter on
{\it Constellation-X}. {\it Chandra} is not sensitive enough to detect
the features. {\it Suzaku} would have been barely sensitive enough to
detect SiXIII and FeXXV, but only along lines of sight fairly far away
from the jet axis, where the jet influence is minimal and therefore it
is not possible to learn much about the interaction dynamics between
the AGN and the ICM. {\it Constellation-X} will have a vastly superior
sensitivity and is therefore much more suitable for that task. In our
simulations, {\it Constellation-X} readily detects these features, can
detect double-peaked FeXXV lines resulting from the expanding shocked
ICM shell, and can detect the re-collapse of the inner ICM after the
AGN has turned off.  We suggest that such observations may be a
powerful new diagnostic of the dynamics of radio-galaxy/ICM
interactions.

\section{Acknowledgments}

We thank Tim Kallman, Barry McKernan and John Vernaleo for stimulating
conversations throughout the course of this work.  We also thank the
referee, Professor Claude Canizares, for a thorough and thoughtful
critique of the original manuscript.  CSR gratefully acknowledges
support from the National Science Foundation under grant AST0205990
and from the {\it Chandra} Cycle-5 Theory \& Modelling program under
grant TM4-5007X.

\label{lastpage}

\end{document}